# Misfit-Dislocation-Mediated Heteroepitaxial Island Diffusion

A. W. Signor*, Henry H. Wu, and Dallas R. Trinkle

*Department of Materials Science and Engineering,*

*University of Illinois, Urbana-Champaign*

## Abstract

Scanning tunneling microscopy combined with molecular dynamics simulations reveal a dislocation-mediated island diffusion mechanism for Cu on Ag(111), a highly mismatched system. Cluster motion is tracked with atomic precision at multiple temperatures and diffusion barriers and prefactors are determined from direct measurements of hop rates. The non-monotonic size dependence of the diffusion barrier is in good agreement with simulations and can lead to enhanced mass transport upon coarsening, in surprising contrast to the traditional island diffusion models where diffusivity reduces with cluster size.

Dislocations are key in the mechanical properties of solids by enabling crystalline materials to deform plastically when subjected to stress orders of magnitude lower than their theoretical critical shear stress [1]. They have also been shown to relieve stress in strained films, greatly affecting the growth mode [2-4], and have been predicted [5,6] but never experimentally implicated in thermally activated surface diffusion of atom clusters. Instead, experimental studies of island diffusion have been mostly limited to homoepitaxial systems where island motion is a result of atomic diffusion events at steps [7-11], with collective cluster motion being negligible for groups of atoms of more than a few, with rare exceptions [12-14]. This work is at the confluence of bulk and surface phenomena by demonstrating that dislocations allow the movement of clusters by reducing the barrier for motion in the same way that they reduce the yield stress of bulk materials: by enabling slip to occur in a piecewise fashion.

Using scanning tunneling microscopy (STM) and molecular dynamics (MD) simulations, we reveal a dislocation-mediated island diffusion mechanism for Cu on Ag(111). This mechanism is active at 80-90 K while edge diffusion and exchange of atoms between clusters is negligible, thus cluster coalescence is the kinetically-preferred coarsening mechanism. Simulations show that the lattice mismatch of ~12% favors dislocation nucleation in islands larger than tetramers with the surprising result that the diffusion barrier for decamers is significantly lower than that for heptamers and is comparable to that for trimers. This non-trivial size dependence is manifest in experiments where clusters containing up to 26 diffuse much faster than smaller clusters.

Experimental measurements were carried out in an Omicron low-temperature STM that can image at 4.5-300 K. The sample stage was enclosed by a cryogenically-cooled metal shroud which kept the the temperature regulated and the sample clean. The Ag(111) substrate was a film grown in an adjoining preparation chamber by evaporating Ag onto Si(111)-7x7 at ~20 K and annealing at ~500 K for 1-2 hours. Films prepared in this manner contain large, flat regions free of defects or steps, ideal for studies of cluster formation and diffusion [15]. To test surface quality, diffusion barriers for atoms and dimers on these substrates were measured and are in good agreement with the published values of 65 and 73 meV, respectively [16]. Once the substrate was prepared, the sample was transferred to the STM stage, cooled to 5 K, and Cu atoms were deposited. The Cu density was determined by counting from images collected at 5 K. The STM stage was warmed to allow atom and dimer diffusion and cluster growth. Cluster sizes were estimated by measuring their area at $X$% of the cluster height, where $X$ was adjusted until the total area agreed with the known adatom density.

The region of interest was imaged every 4.26 min for ~80 hours at each temperature, 80, 83, 85, and 87 K. This enabled diffusion barriers and prefactors to be measured directly for individual clusters. Fig. 1 shows a sequence of STM images collected at 80 K. What is surprising is that the 13-atom cluster underwent significant diffusion, while the 7- and 15-atom clusters were immobile. This suggests a magic size effect where diffusion barriers are reduced for islands of particular sizes. Note that there was no change in size or shape, ruling out edge diffusion or atom exchange between islands. The dotted line guides the eye along the diffusion path of the 13-atom cluster. The one-dimensional path of motion along the close-packed $[0\bar{1}1]$ direction is significant because fcc metals slip on {111} planes in [110] directions. This suggests that the active diffusion mechanism is likely related to dislocation glide.

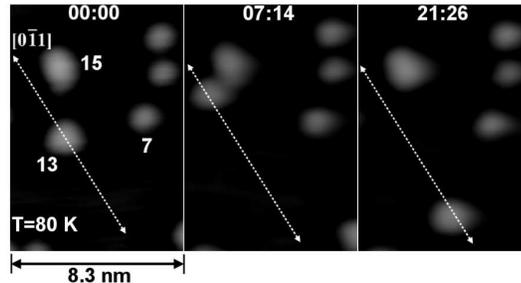

FIG. 1: STM images (-200 mV, 0.5 nA) at 80 K showing diffusion of a 13-atom Cu cluster on Ag(111). The relative times of the images are given in h:min. The arrow indicates a diffusion trajectory along $[0\bar{1}1]$. This cluster undergoes significant diffusion, while the neighboring 7- and 15-atom clusters are immobile.

From STM movies, using a particle tracking program [17], we determine trajectories with atomic precision, as shown in Fig. 2 for a 10-atom cluster at 85 K. The positions occupied by the cluster centroid are plotted vertically on the left, revealing a discrete set of adsorption sites separated by the Ag nearest-neighbor distance. This shows that the cluster hops collectively between equivalent sites. If diffusion occurred through independent displacements of periphery atoms, hop lengths for the centroid would be a fraction of this distance and inversely proportional to island size. Diffusion measurements for the cluster in Fig. 2 at 83, 85, and 87 K showed that it visited 9, 12, and 20 sites after 50, 120, and 230 hops, respectively. This demonstrates that the diffusion is close to an unrestricted 1-D random walk since the number of sites visited in a walk of $n$ hops is expected to be $(8n/\pi)^{1/2}$, which yields 11, 17, and 24 sites for the cases above [18].

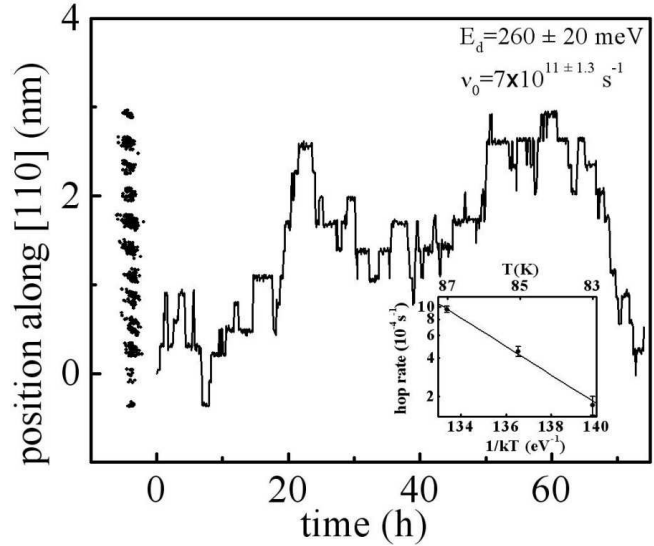

FIG. 2: Diffusion trajectory for a 10 atom cluster doing a 1-D walk along [110] at 85 K. Each point on the left represents the cluster's centroid from a data set containing ~1100 frames. The plot as a function of time makes it possible to determine the mean hop rate. In the inset, the temperature dependence of the hop rate yields an activation energy of 260 ± 20 meV, and an attempt frequency of $7 \times 10^{11 \pm 1.3}$ s$^{-1}$.

The plot of position vs. time in Fig. 2 shows that the mean time between hops is long compared to the frame rate, ensuring that all hops are counted and mean hop rates can be measured directly. The mean hop rate for this cluster at 83, 85, and 87 K is plotted against $1/kT$ in the inset, giving an activation energy of $260 \pm 20$ meV and a prefactor of $7 \times 10^{11 \pm 1.3}$ s$^{-1}$. Several 10-atom clusters were followed at multiple temperatures and all had barriers within a standard error of 260 meV and the prefactors were within a standard error of $10^{12}$ s$^{-1}$, which is the usual attempt frequency for thermally activated processes.

The experiments show that low diffusion barriers are not limited to clusters larger than heptamers. Fig. 3 is a series of STM images showing the diffusion of a pentamer at 80 K. In this case, motion was too fast for direct hop rate measurement, but the number of distinct sites visited was determined to be 55. By assuming an unrestricted random walk with a prefactor of $10^{12}$ s$^{-1}$, the barrier was estimated to be ~210 meV. Similar measurements for 13-, 14- and 26-atom clusters at 83 K give barriers of 225, 240 and 250 meV, respectively, showing that this mechanism is viable beyond the decamer. In comparison, these diffusion barriers are much lower than the barriers for homoepitaxial island migration through edge diffusion on either Cu(111) or Ag(111), ~500 meV [9]. The non-monotonic size dependence of the diffusion barrier is clearly more complex than the simple single magic size effect predicted by Hamilton [6] for a mismatched system.

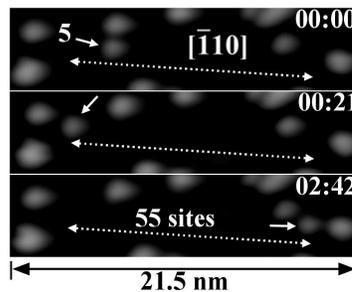

FIG. 3: STM images (-200 mV, 0.5 nA) showing diffusion of a 5-atom cluster at 80 K. The relative times of the images are given in h:min. While pentamer diffusion at 80 K was too fast for direct hop rate determination, the barrier was estimated to be ~210 meV, assuming an unrestricted random walk.

MD simulations were conducted to investigate the atomic processes at work. An embedded-atom method potential parameterized for Cu/Ag(111) [19] determines the diffusion barriers and mechanisms for selected islands. This potential slightly overestimates the monomer and dimer diffusion barriers, giving 93 and 88 meV, respectively. The potential is optimized to produce accurate island geometries, energies, and kinetics. High-temperature annealing allows the equilibrium island shapes to be determined. Molecular dynamics simulations and dimer method [20] search the phase-space for possible diffusion transitions, and the nudged-elastic band method [21] determines the energy barriers and the atomic-scale mechanisms.

Fig. 4 summarizes the MD results for 3-, 7-, 5-, and 10-atom clusters, which show that islands of certain sizes and shapes allow metastable dislocations, leading to reduced diffusion barriers. The trimer in Fig. 4 (a) moves through simultaneous glide with a barrier of 287 meV, roughly three times the simulated monomer diffusion barrier. The heptamer (b) also moves in a collective fashion, though the atoms do not cross bridge sites simultaneously. In the transition state, the top right portion of the island moves towards hcp stacking prior to the lower left, reducing the barrier to 490 meV, about five times the monomer barrier.

Fig. 4 (c) and (d) show that a different, low-barrier mechanism involving a metastable dislocation (dotted lines) is accessible to the pentamer and decamer. Surprisingly, the diffusion barrier for the decamer, 283 meV, is little more than half that of the heptamer and even slightly lower than the trimer barrier. The pentamer and decamer diffusion barriers, 214 and 283 meV, respectively, are in excellent agreement with the experimental values of ~210 and 260±20 meV. The overestimation of barriers in the simulations is expected, based on the monomer and dimer simulation barriers. The diffusion process for the decamer in Fig. 4 (d) proceeds as follows. Starting from a configuration in which all Cu atoms are in fcc sites, F10, a metastable state with 5 atoms in fcc sites and 5 in hcp sites, F5H5, is accessed. The dashed line indicates a dislocation with Burgers vector $\vec{b} = 1/6[\bar{2}11]$ separating the fcc and hcp regions. If the remaining fcc atoms follow to H10, the center of mass is displaced by one Burgers vector. A second dislocation with $\vec{b} = 1/6[12\bar{1}]$ can nucleate in H10, and the result is net displacement by $1/2[\bar{1}10]$. Symmetry allows F10 to accommodate dislocations with $\vec{b} = 1/6[\bar{2}11]$ or $\vec{b} = 1/6[1\bar{2}1]$ while H10 can accommodate $\vec{b} = 1/6[\bar{1}2\bar{1}]$ or $\vec{b} = 1/6[2\bar{1}\bar{1}]$. Thus, successive dislocation events allow for reptation [22,23] with forward, backward, or zero net displacement along [$\bar{1}10$] with equal probability and a barrier of 283 meV for a complete fcc-fcc step. The short lifetimes of H10 (~$10^{-4}$ s) and H5F5 (~$10^{-8}$ s) compared to F10 (on the order of seconds) would prevent them from being observed with STM and any image of the cluster would show it in the F10 configuration. The pentamer (c) moves in an analogous manner. Despite the 3-fold symmetry of fcc {111}, the vast majority of clusters in experiments diffuse in 1D. This suggests that there is a correlation between the directions of motion of each half of the island. There is the possibility that cluster interactions break symmetry leading to the correlation, but the 1D walks persist over long distances despite the changing local environment. This indicates that the 1D walks are an intrinsic property of the clusters.

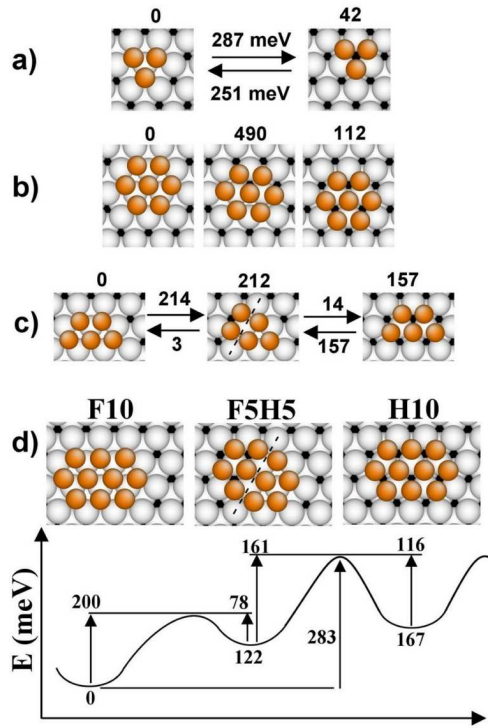

FIG. 4: Molecular dynamics simulations of 3-, 7-, 5-, and 10-atom Cu clusters on Ag(111). The trimer (a) migrates via simultaneous glide, with all atoms moving across bridge sites at the same time. The heptamer (b) moves in a dislocation-like mechanism, where the transition state contains atoms in both fcc and hcp sites. The pentamer (c) and decamer (d) diffuse via a different misfit dislocation mechanism, where the states containing the dislocation are metastable and the dislocation lines are oriented along [110]. All energy values are given in meV.

More extensive simulations show that the low-barrier reptation mechanism, as shown for the decamer and pentamer in Fig. 4, is inaccessible to closed-shell structures like the trimer and heptamer. To preserve Cu-Cu bonds, the dislocations in this process must nucleate between close-packed rows of the same length and with Burger's vectors that bring the atoms closer together [24]. This is why the only dislocations allowed in the F10 or H10 configurations are $1/6[\bar{2}11], 1/6[1\bar{2}1]$ or their directional opposites. This rule means that besides island size, shape is important—something not considered in previous simulations of dislocation mechanisms. There is indication in the experimental measurements that shape can be as important as size because some initially-immobile clusters are spontaneously mobilized subsequent to a shape change while maintaining their size.

Our diffusion model is further supported by the atomic resolution image in Fig. 5 collected at 100 K after a 90-second anneal to ~200 K which allowed for significant coarsening. Despite this, some heptamers survived the anneal and were immobile at 100 K while all other islands, many bi-layer, contained 20 or more atoms. This is consistent with the simulations, which predict the heptamer to have the highest diffusion barrier among clusters containing up to 14 atoms, as its size and shape prevent it from accessing a low-energy, metastable dislocation state.

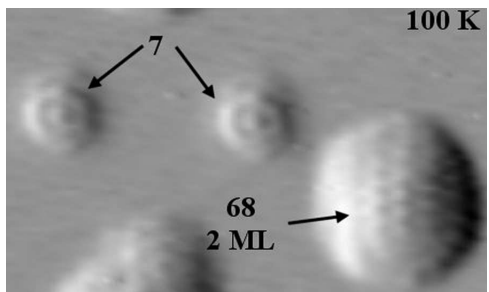

FIG. 5: Atomic resolution image at 100 K after a ~90 sec anneal at ~200 K. Heptamers survived the anneal and were immobile at 100 K. The other clusters on the surface contained 20 or more atoms, having formed from cluster-cluster coalescence. Many bi-layer islands were present, like the 68-atom island shown.

We have shown that a dislocation mechanism provides a pathway to coarsening through cluster diffusion for clusters as large as 26 atoms with barriers significantly lower than those for edge diffusion or Ostwald ripening. This mechanism leads to significantly reduced diffusion barriers for islands of the proper sizes and shapes that favor metastable dislocations with the surprising result that large islands can move more easily than smaller ones. Thus, in much the same way that dislocations reduce the yield stress of bulk metals from their theoretical values, they also reduce island diffusion barriers. It is clear that this mechanism is promoted by lattice mismatch, which reduces the energy cost of bringing the Cu atoms closer together, and it is likely a general phenomenon applicable to similarly mismatched systems.

We thank John Weaver, Robert Butera, Parasuraman Swaminathan, Celso Aldao, and Gert Ehrlich for valuable discussions. This work is supported by NSF/DMR grant 0703995.

---